\documentclass[fleqn,twoside]{article}
\usepackage{espcrc2}

\usepackage{graphicx}

\newcommand{\AmS}{{\protect\the\textfont2
  A\kern-.1667em\lower.5ex\hbox{M}\kern-.125emS}}

\title{Short time-scale X-ray variability of Active Galactic Nuclei}

\author{Alexander V. Halevin\address[ONU]{Department of Astronomy, Odessa National 
University, \\
T.G.Shevchenko park, 65014, Odessa, Ukraine,\\ halevin@astronomy.org.ua}%
}
       
\begin{document}

\begin{abstract}
In this work we have investigated short time-scale variability of 
NGC 4051 and NGC 4388. The quasi-steady QPO's time scale of 
3920$\pm$17 sec was detected for NGC 4051. For the case of NGC 
4388 we investigated rapid variability of the parameters of 
spectral models. To obtain these results, a new ``Moving Fit" 
method for processing of the trailed X-ray spectrograms was 
developed. \vspace{1pc} 
\end{abstract}

\maketitle

\section{Introduction}

Short time-scale X-ray variability from dozen of seconds to hours 
in AGNs is explained as a result of different kinds of processes, 
which happen close to the central engine. The possible origins are 
changes of the accretion rate, flares in accretion disk, motion of 
the hot spots around black hole and, for longer time-scales, the 
motion of hydrogen clouds, which obscure the central source. 

Here we present results of investigations of short time-scale 
variability of active galactic nuclei using as an example Chandra 
observations of two bright Seyfert type galaxies NGC~4051 and 
NGC~4388. Both archive data sets (3144 and 2983) respectively were 
processed using CIAO version 2.3 package. 

\section{Flux variability of NGC 4051}

Chandra observations of NGC~4051 were made at HRC LETG mode with 
exposure of approximately 92 ksec on January 01, 2002. To study 
fast flux variability we extracted only zeroth order image of 
NGC~4051. For a source and background extracting and making of the 
light curve we have used \it dmextract \rm utility of CIAO 
package. 

\begin{figure}[h]
\vspace{0.1cm} \hspace{0cm} 
\resizebox{\hsize}{!}{\includegraphics{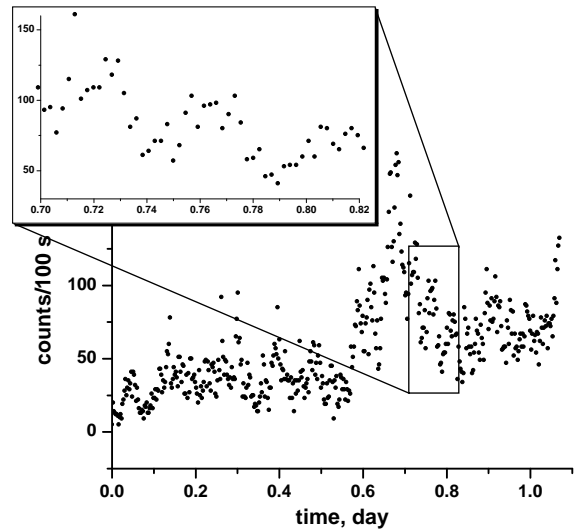}} \caption{The 
light curve of NGC~4051 obtained with Chandra HRC detector on 
January 01, 2002 (observation identification number 3144).} 
\label{fig01} \vspace{0cm} 
\end{figure}
The resulting light curve is shown in the Fig.1. During these 
observations, NGC~4051 was very active with very strong high 
amplitude variability. In order to investigate a character of 
variability we have used wavelet method, which is excellent for 
detection of the non-coherent or weakly coherent variations. Many 
papers are devoted to an application of the wavelet method to 
periodic or multiperiodic processes 
(\cite{Gou91,Sza94,Fri98,Hal02}). The long-term variations have 
\begin{figure}[h]
\vspace{0cm} \hspace{0cm} 
\resizebox{\hsize}{!}{\includegraphics{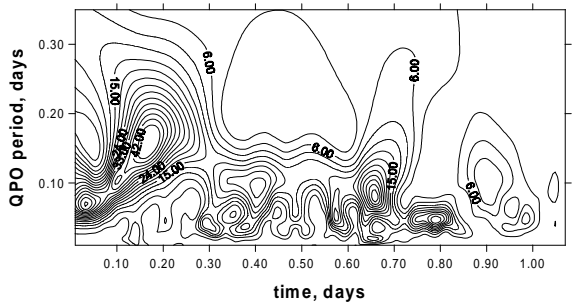}} \caption{Wavelet 
map for detrended light curve of NGC~4051.} \label{fig2} 
\vspace{0cm} 
\end{figure}
\begin{figure}[h]
\vspace{0.5cm} \hspace{0cm} 
\resizebox{\hsize}{!}{\includegraphics{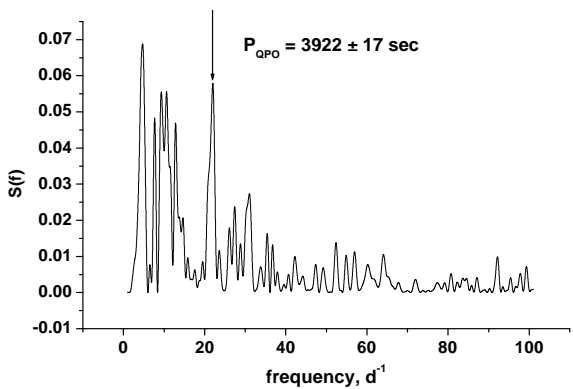}} \caption{The 
power spectrum for NGC~4051.} \label{fig3} \vspace{0cm} 
\end{figure}
been smoothed by using the method of running parabola 
(\cite{And97}). The optimal value of the filter half--width 
$\Delta t=0.217~d$ has been determined from maximization of the 
``signal-to-noise" ratio. To avoid apparent effects of 
low--frequency trends on the test--functions at high frequencies, 
the original data have been detrended, i.e. the running parabola 
fit was subtracted from the observations. For these detrended time 
series, the test functions have been computed using the code 
described by \cite{And94}. For visualization, we have used the 
Weighted Wavelet Z-transform (WWZ) test-function \cite{Fos96}, 
which is characterized as having the best contrast among other 
test functions. 

\begin{figure}[h]
\vspace{0.7cm} \hspace{0cm} 
\resizebox{\hsize}{!}{\includegraphics{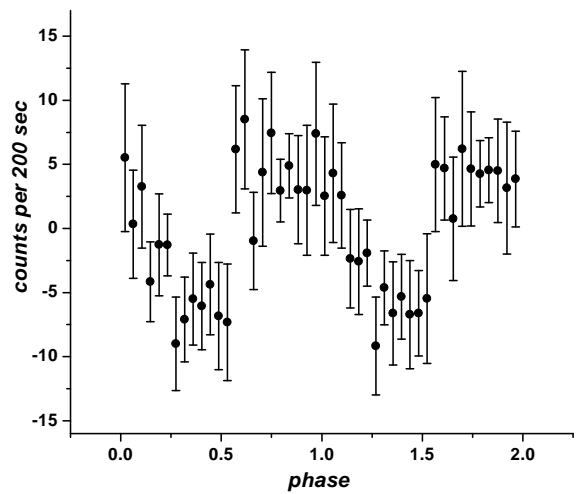}} \caption{The 
phase folded light curve for 
$P_{QPO}~=~0.04550\pm0.00014~d~=~3922\pm17~sec$.} \label{fig4} 
\vspace{0cm} 
\end{figure}

The resulting wavelet map one can see in the Fig. 2. It is clear 
an existence of the QPOs with different time-scales. The QPOs are 
coexistent with flares and it is very difficult to distinguish 
them. One can see that after powerful outburst QPOs almost
disappeared. 

To check the presence of the typical frequency we have computed 
the power spectrum (using the method of Fourier analysis, which 
was described by \cite{Dee75}. The power spectrum one can see in 
the Fig.3. There is very significant peak which corresponds to 
period 3922 $\pm$ 17 sec. The phase folded curve for this period 
one can see in the Fig.4. 

To check the existence of the same frequency in another runs, we 
have used the archival observations of Chandra obtained with ACIS 
detector in April 2000 (archive number 859, \cite{Coll01}). In the 
power spectrum for this run we have found also significant but 
less pronounced peak 3950 $\pm$ 68 sec. Probably, this time-scale 
corresponds to the orbital period of the inner parts of accretion 
disk. Assuming the distance of about 10 $R_{g}$ we can estimate 
black hole mass as $1.4 \times 10^{6} M_{\odot}$. 

\section{Investigations of the spectral variability of NGC~4388 using ``Moving Fit" method}

Rapidly growing sensitivity of X-ray detectors allows to 
investigate the spectral variability with relatively short 
time-scales. In order to improve our capabilities in this area we 
have made the package of the programs, which performed for 
analysis of the trailed X-ray spectra (Fig.5). This package 
includes some capabilities of XSPEC and XRONOS packages. It was 
written using C++ language with CFITSIO and PGPLOT libraries. In 
the future we are going to make C++ module plus Tcl script for 
XSPEC for processing of the trailed X-ray spectra. 

\begin{figure}[h]
\vspace{0cm} \hspace{0cm} 
\resizebox{\hsize}{!}{\includegraphics{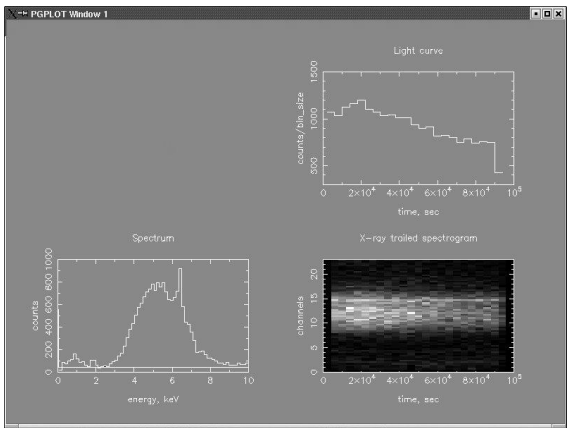}} 
\caption{Interface of the program for processing of the trailed 
spectra.} \label{fig5} \vspace{0cm} 
\end{figure}

In the present version of the package as an input parameter were 
chosen binned both in time and energy scale data presented as an 
FITS image, which we can obtain using \it dmextract \rm utility in 
the case of Chandra observations. A detector response matrix file 
(``rmf") and an auxilliary response file (``arf") are also 
necessary. The present version of our package makes spectral fit 
with limited number of models like as power law, absorption 
cut-off, Compton reflection component, gaussian-profile emission lines and some elementary 
mathematical functions. For minimizing of the $\chi^{2}$ 
statistics a genetic algorithm-based optimization was performed. 

\begin{figure}[h]
\vspace{0.8cm} \hspace{0cm} 
\resizebox{\hsize}{!}{\includegraphics{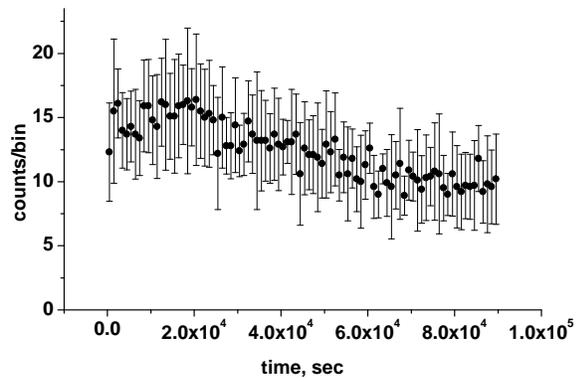}} \caption{Light 
curve for Chandra ACIS observations of NGC 4388.} \label{fig6} 
\vspace{0cm} 
\end{figure}

The primary idea of ``Moving Fit" method is building of the models 
for a located in time slice of the spectral data and the data from 
the neighbor time bins taken with some additional weights (the 
implementation of the ``weighted moving average" method for each 
spectral channel). It is possible to use of different kinds of 
kernels like as simple rectangular box, gaussian, etc. 

Our package also has additional capabilities for complicated 2D 
fitting, for example, fitting of the emission line behaviour by 
sine or any other function. 

In the present work we have used this package for investigations 
of the spectral variability of obscured Seyfert type galaxy 
NGC~4388. Observations (dataset 2983) were obtained on March 6, 
2002 with Chandra ACIS detector in the LETG mode. Light curve for 
this AGN one can see in the Fig.6. Time\&energy binned image and 
``rmf" and ``arf" files were created with \it dmextract \rm 
utility and \it psextract \rm script of CIAO package only for the 
zeroth order image. 

Time bin was chosen as 4 ksec. The resulting low energy resolution 
image was fitted with power law, gaussian $Fe~K\alpha$ line near 
6.4 keV and photoelectric absorption calculated following 
\cite{Mor83}. A power law photo-index was fixed equal to $1.92 \pm 
0.11$ as we have derived for full data set using the same set of 
models in XSPEC package. 

As a smoothing function we have used simple rectangular box with 
half-width 10 ksec. Resulting curves for hydrogen column, power 
low normalization index and $Fe~K\alpha$ parameters are presented 
in the Fig.7. The error bars are 99 \% confidence intervals. 

\begin{figure}[t]
\vspace{0cm} \hspace{0cm} 
\resizebox{\hsize}{!}{\includegraphics{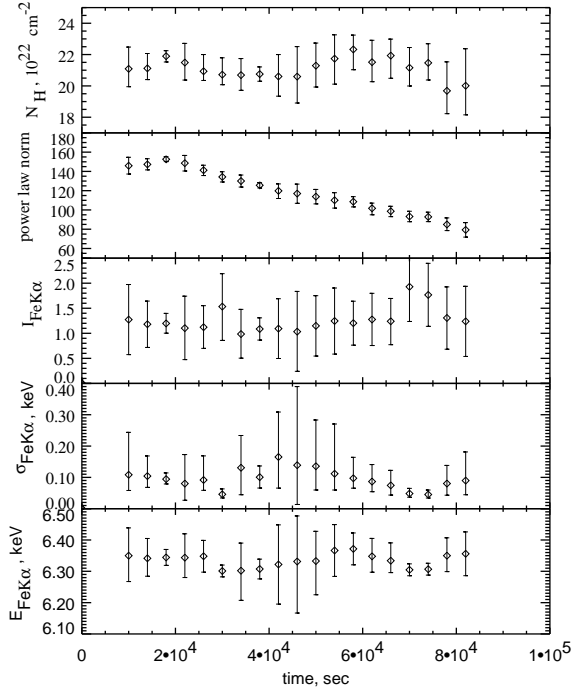}} 
\caption{Variability of the NGC~4388 spectrum model parameters 
with time, derived using ``Moving Fit" method.} \label{fig7} 
\vspace{0cm} 
\end{figure}

First of all the curve for power law normalization parameter is
practically identical to the light curve (Fig.6). There is also 
some possibility for smooth changes of the hydrogen column with 
time-scale about 40 ksec. The curve for the $Fe~K\alpha$ line 
central intensity shows flare during which $\sigma_{K\alpha}$ and 
$E_{K\alpha}$ were probably suppressed. 

So you can see, that this method allows to investigate rapid X-ray 
spectral variability and could be implemented for different kinds 
of objects. \\ 


{\it Acknowledgements. \rm Author is thankful to prof. Ivan 
Andronov for helpful discussions and to Alina Streblyanskaya, who 
stimulated an appearance of this work.\\}

\end{document}